%% file: main.tex
\RequirePackage{amsmath}
\documentclass[runningheads]{llncs}
\usepackage[T1]{fontenc}
\usepackage{graphicx}
\usepackage{csquotes}
\usepackage[table]{xcolor}
\usepackage{tabularx}
\usepackage{orcidlink}
\usepackage{adjustbox} 
\usepackage{graphicx,multirow} 
\usepackage{float}
\usepackage{ccicons}

\usepackage[inkscapeformat=eps,inkscape=false]{svg}
\svgpath{./}
\usepackage{dirtytalk} 
\input{acrons}
\input{listings-style}

\usepackage{multicol} 

\usepackage{pgf-umlsd}
\usetikzlibrary{positioning}
\usetikzlibrary{patterns}
\usetikzlibrary{calc}
\usepackage{booktabs} 
\usepackage[most]{tcolorbox} 
\usepackage{enumerate}
\usepackage{xurl}
\usepackage{hyperref}      

\usepackage[simplified]{pgf-umlcd} 
\usepackage[normalem]{ulem} 

\input{table-markers}

\usepackage{enumitem} 

\begin{document}
\title{%
    The Emperor is Now Clothed: \\ A Secure Governance Framework for Web User Authentication through Password Managers%
    \thanks{%
        This is the ePrint of a paper to appear at the proceedings of ICICS~2024. 
        \\ 
        \includegraphics[height=1.8ex]{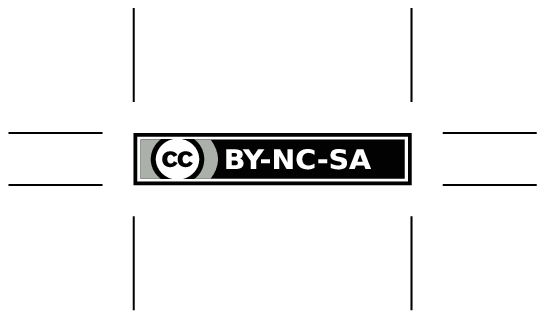}
        This work is licensed under \href{https://creativecommons.org/licenses/by-nc-sa/4.0/}{CC~BY-NC-SA~4.0}. 
    }%
}
\titlerunning{A Secure Password Manager Governance Framework}
%
\author{Ali Cherry\orcidlink{0000-0002-4111-3103} \and
Konstantinos Barmpis\orcidlink{0000-0002-0864-0956} \and
Siamak F.\ Shahandashti\orcidlink{0000-0002-5284-6847}}
\authorrunning{A. Cherry et al.}

\institute{University of York, York, United Kingdom
\email{\{ali.cherry,konstantinos.barmpis,siamak.shahandashti\}@york.ac.uk}}

\maketitle              

\begin{abstract} 
Existing approaches to facilitate the interaction between password managers and web applications fall short of providing adequate functionality and mitigation strategies against prominent attacks. 
HTML Autofill is not sufficiently expressive, 
Credential Management API does not support browser extension password managers,
and other proposed solutions do not conform to established user mental models.
In this paper, we propose \protoname, a browser-based governance framework that mediates the interaction between password managers and web applications. 
Two APIs are designed to support \protoname\ acting as an orchestrator between password managers and web applications. 
An implementation of the framework in Firefox is developed that fully supports registration and authentication processes. 
As an orchestrator, \protoname\ is able to authenticate web applications and facilitate authenticated key exchange between web applications and password managers, which as we show, can provide effective mitigation strategies against phishing, cross-site scripting, inline code injection (e.g., by a malicious browser extension), and TLS proxy in the middle attacks, whereas existing mitigation strategies such as Content Security Policy and credential tokenisation are only partially effective. 
The framework design also provides desirable functional properties such as support for multi-step, multi-factor, and custom authentication schemes.
We provide a comprehensive security and functionality evaluation and discuss possible future directions.

\end{abstract}

\input{sections/intro/v2}

\input{sections/background/v2}

\input{sections/method/v2}

\input{sections/security/v2}

\section{Functionality Evaluation}
\label{sec:fun-eval}
Deployability is discussed as a crucial property of security-enhancing technologies~\cite{bonneau_quest_2012} and a significant factor for secret manager adoption~\cite{pearman_why_2019,bonneau_quest_2012}.
A framework that is designed to orchestrate the relationship between secret managers and web apps hence need to consider the developers on both sides. 
In this section, we discuss this aspect and evaluate \protoname\ in terms of functionalities it provides for secret managers and web apps.  
Comparison between existing frameworks and \protoname\ with respect to these functionalities is presented in Table~\ref{table:functionality_and_deployability_overview}.


\begin{table}[t!]
    \newcommand{\seprow}{\\}
    \newcommand{\notsuppbutcan}{\emptycircle{}}
    \newcommand{\supp}{\cmark{}}
    \newcommand{\notposs}{\xxmark{}}

    \setlength{\tabcolsep}{3pt}
    \centering
        
    \caption{Functional capability comparison between existing frameworks and \protoname.}
    \label{table:functionality_and_deployability_overview}
    \begin{tabular}{lcccc} 
        \toprule
        Functional Capability & HTML Autofill  & Cred Mgmt API & ByPass & \protoname{} \\
        \midrule
        Multi-Step \& Multi-Factor Auth
            & \notposs & \notsuppbutcan & \notsuppbutcan & \supp
            \seprow
        Developer Control over UX
            & \hmark & \supp & \notposs & \supp
            \seprow
        Conventional Usage Pattern
            & \supp & \supp & \notposs & \supp 
            \seprow
        \scm{} Extensions
            & \supp & \notsuppbutcan & \supp & \supp 
            \seprow
        Flexible Account Design
            & \notsuppbutcan & \notsuppbutcan & \notsuppbutcan & \supp
            \seprow
        Custom Authentication
            & \notposs & \notsuppbutcan & \notsuppbutcan & \supp
            \seprow
        \scm{} Allowlisting
            & \notposs & \notsuppbutcan & \notsuppbutcan & \supp
            \seprow
        \bottomrule
    \end{tabular}
    \vskip 0.25em
    \notposs: not possible, 
    \notsuppbutcan\!: not currently supported, 
    \hmark\!: partially supported, 
    \supp\!: supported
    \hfill
\end{table}

\paragraph{Support for Multi-Step and Multi-Factor Authentication.}
Existing frameworks only support \say{one-shot} scenarios, where all credentials are expected to be transmitted in one instruction.
\protoname\ models provision of multiple authentication credentials as successive \emph{challenges}, where each challenge is responded to in a series of instructions. 
This allows flexible support for multi-step and multi-factor authentication protocols, including multi-round protocols such as SRP.

\paragraph{Developer Control over UX.}
\protoname\ leaves the web app and secret manager developers in charge of UX and does not enforce any requirements on their UIs. 
The automation is to a large extent transparent to the user and occurs at the script level. 
This is in contrast with ByPass which does away with the web app UI and enforces a predetermined UX. 
Although web app developers have control over the login UI in HTML Autofill, they cannot control when the credential transfer is conducted, and hence they only have partial control over the UX.

\paragraph{Maintaining the Conventional Usage Pattern.}
A typical user expects to visit a web page on their browser to log in to the corresponding web app. 
Client-side frameworks align with this mental model, whereas more radical frameworks like ByPass 
expect users to conduct all their account-related tasks with web apps through the secret manager's interface. 

\paragraph{Support for Secret Manager Extensions.}
\protoname\ is designed to work with both native and third-party secret managers that are installed as extensions. 
This, in turn, supports choice for end users and encourages an open ecosystem. 
Credential Management API however is designed for native secret managers. 

\paragraph{Flexibility over Account Design.}
\protoname\ allows web apps to define their account structure as any combination of the supported fields.
Besides, \protoname{} has built-in support for categorisation which allows the possibility of multiple role-based 
accounts for the same user on the same web app. 

\paragraph{Support for Custom Authentication.}
\protoname\ allows for custom authentication challenges to be initiated following approval from the selected secret manager. 
These challenges are in the form of custom messages validated using an app-specified JSON schema, and allow flexibility in extending the supported authentication mechanisms.  
Existing frameworks do not provide such support.

\paragraph{Support for Secret Manager Allowlisting.}
Web apps with strict security policies may wish to restrict the list of secret managers they trust with user credentials. 
\protoname\ provides the facility for web apps to specify an allowlist of approved secret managers. 
Such a facility is not provided by any of the existing frameworks.

\section{Conclusions and Future work}
\label{sec:fw}

We presented \protoname{}, a novel framework for web-based authentication, which supports the registration and authentication processes, including identity information management and external identity (e.g., phone or email) attestation. 
It supports extension secret manager as well as native user agents, thereby maintaining ecosystem openness for users of any browser-based secret manager.

Our security evaluation demonstrated that \protoname\ provides valuable security services which mitigate against prominent attacks. 
Notable among these services is the provision of end-to-end encrypted channels between secret managers and web app back-ends which render credential theft attacks via code injection ineffective, even when they are carried out by entities with privileged access such as browser extensions and TLS proxies.

Through our functionality evaluation, we showed that \protoname\ is the only platform of its kind that provides programmatic access to web apps and secret managers while supporting both native and extension secret managers and preserving control over user experience for developers and usage mental models for users. 
Flexible account design and support for multi-step, multi-factor, and custom authentication protocols are among other benefits provided. 

While it is expected that such a solution will have some integration effort required both from web applications and secret managers, \protoname{} tries to minimise such effort, keeping various services requiring additional effort, e.g., web app authentication and authenticated key exchange, entirely optional. 
A detailed analysis of the integration effort can be found in~\cite[Section 4.2.2]{cherry_secure_2024}.
We suggest future work to conduct studies with developers and end users to investigate the usability of \protoname\ integration and utilisation in more detail.



%
%
%
%



\bibliographystyle{splncs04}

\input{main.bbl}
\appendix

\section{Proof of Concept Implementation}
\label{sec:implementation}

We briefly discuss the proof of concept implementation in this section. 
All project artefacts are available at \url{https://github.com/alichry/berytus}. 
Build instructions are provided as well as binaries 
for our extended Firefox browser.

\begin{figure}
    \centering
    \begin{minipage}{0.49\textwidth}
        \includegraphics[width=\linewidth]{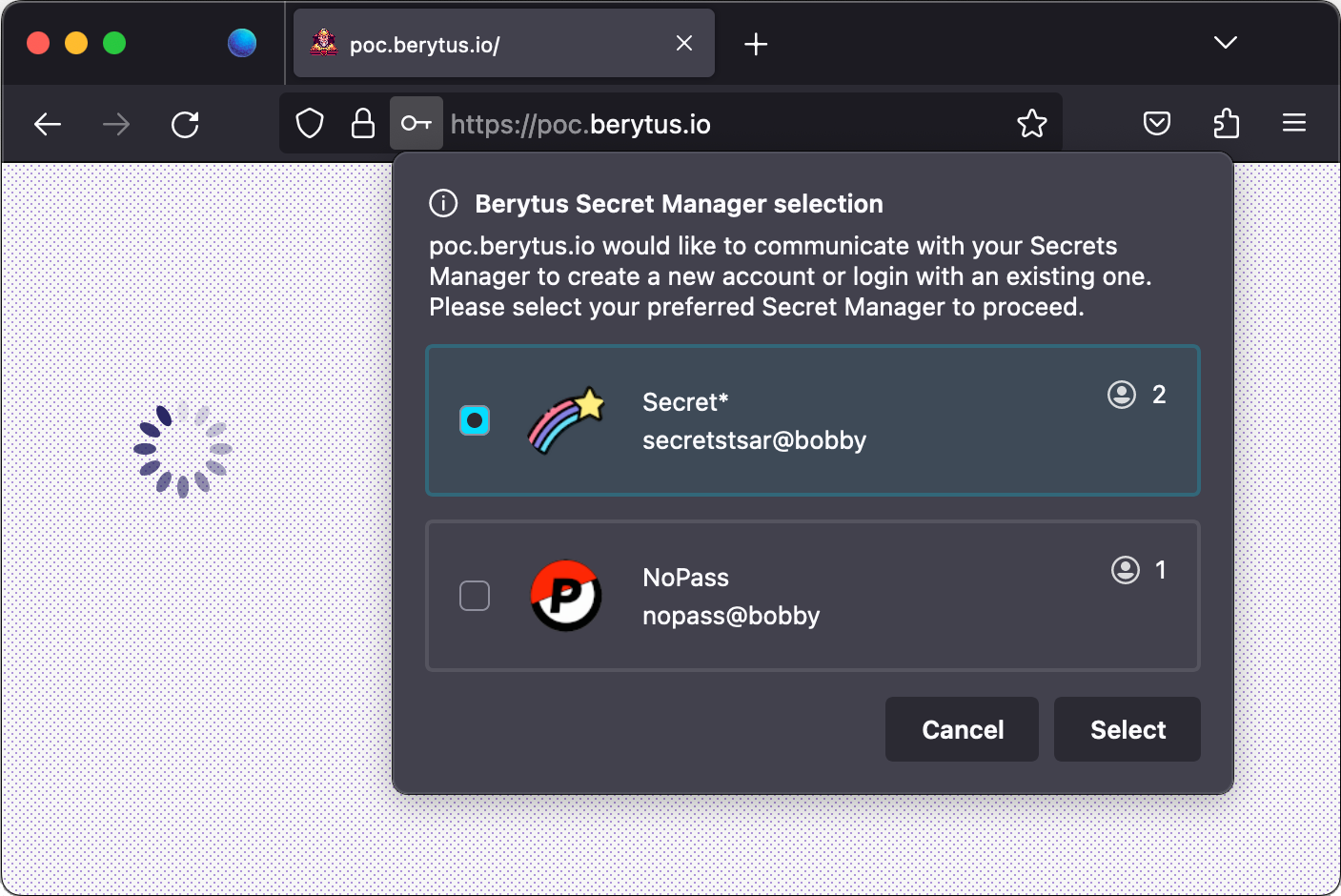}
    \end{minipage}
    \begin{minipage}{0.49\textwidth}
        \includegraphics[width=\linewidth]{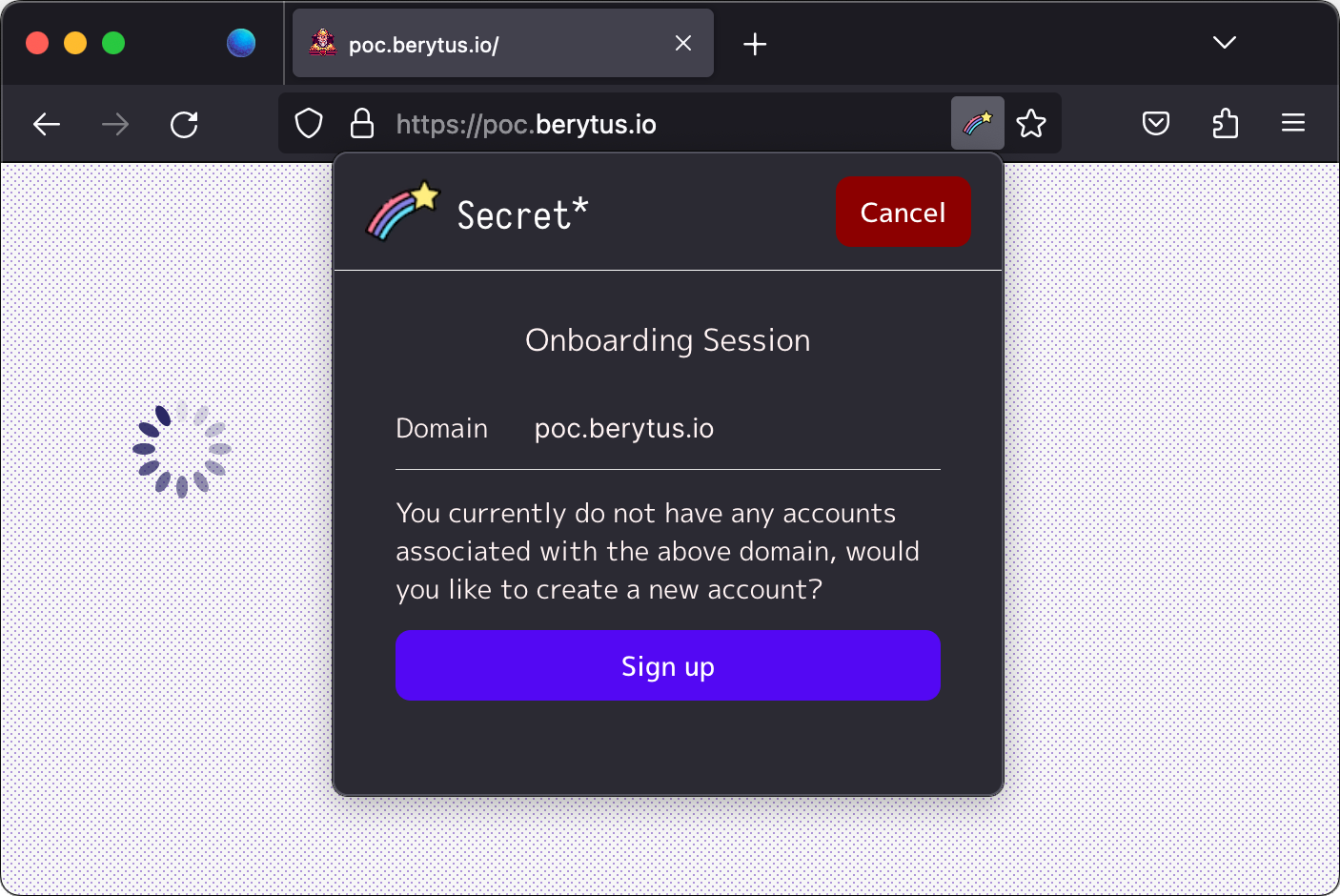}
    \end{minipage}    
    \caption{
        Left: \protoname\ secret manager selection prompt.
        Each listed secret manager displays the number of registered accounts associated with the web app. 
        Right: Secret* login operation approval prompt when the user does not have any registered accounts. 
        In both cases, domain-based credential mapping was used.}
    \label{fig:berytus-scm-prompt}
\end{figure}

\paragraph{API Implementation.}
To demonstrate the feasibility of our framework, we have extended the Mozilla Firefox browser (v116.0a1) to implement the Web API and the WebExtensions API natively. 
These two ends are linked with one another through the \protoname{} Core API. The Core API handles secret manager selection (Figure~\ref{fig:berytus-scm-prompt}~(left)), channel and operation management, inter-process communication and other functional requirements. The Web API is implemented in C++, the WebExtensions API and the Core API in TypeScript/JavaScript.

\paragraph{Secret Manager Implementation.}
We developed \emph{Secret*} (\say{secret star}), a secret manager that integrates with the \protoname{} WebExtensions API.
Secret* has three components: 
(1) The background script, where the request handler is implemented and registered; (2) The user interface facility which hosts the extension pop up pages 
used for user prompts;
(3) The storage facility to store and retrieve data when processing requests.
Figure~\ref{fig:berytus-scm-prompt}~(right) shows how Secret* retrieves the user's intent during a login operation. 

\paragraph{Web Application Implementation.}
We developed a full-stack web application with an authentication subsystem
and proceeded to integrate \protoname{}. This serves as a 
real-world example, demonstrating feasibility and showing the changes
needed on the server-side and client-side to complete the integration.

\end{document}

%% file: acrons.tex
\usepackage{xstring} 
\usepackage{mfirstuc} 

\newcommand{\protoname}{Berytus}

\newcommand{\webapp}{Web App}

\newcommand{\scm}{Secret Manager}

\newcommand{\cryptoactor}{Crypto Actor}
\newcommand{\originactor}{Origin Actor}

\newcommand{\signingkey}{Signing Key}

\newcommand\an[1]{%
  \StrLeft{#1}{1}[\firstchar]%
  \IfStrEq{\firstchar}{a}{an }{%
    \IfStrEq{\firstchar}{A}{an }{%
    \IfStrEq{\firstchar}{o}{an }{%
    \IfStrEq{\firstchar}{O}{an }{%
    \IfStrEq{\firstchar}{u}{an }{%
    \IfStrEq{\firstchar}{U}{an }{%
    \IfStrEq{\firstchar}{i}{an }{%
    \IfStrEq{\firstchar}{I}{an }{a }%
    }%
    }%
    }%
    }%
    }%
    }%
  }%
  #1%
}

%% file: listings-style.tex
\definecolor{lightgray}{rgb}{0.95, 0.95, 0.95}
\definecolor{darkgray}{rgb}{0.4, 0.4, 0.4}
\definecolor{editorGray}{rgb}{0.95, 0.95, 0.95}
\definecolor{editorOcher}{rgb}{1, 0.5, 0} 
\definecolor{editorGreen}{rgb}{0, 0.5, 0} 
\definecolor{orange}{rgb}{1,0.45,0.13}		
\definecolor{olive}{rgb}{0.17,0.59,0.20}
\definecolor{brown}{rgb}{0.69,0.31,0.31}
\definecolor{purple}{rgb}{0.38,0.18,0.81}
\definecolor{lightblue}{rgb}{0.1,0.57,0.7}
\definecolor{lightred}{rgb}{1,0.4,0.5}
\usepackage{upquote}
\usepackage{listings}

\definecolor{mygreen}{rgb}{0,0.6,0}
\definecolor{mygray}{rgb}{0.5,0.5,0.5}
\definecolor{mymauve}{rgb}{0.58,0,0.82}

\lstdefinelanguage{CSS}{
  keywords={color,background-image:,margin,padding,font,weight,display,position,top,left,right,bottom,list,style,border,size,white,space,min,width, transition:, transform:, transition-property, transition-duration, transition-timing-function},	
  sensitive=true,
  morecomment=[l]{//},
  morecomment=[s]{/*}{*/},
  morestring=[b]',
  morestring=[b]",
  alsoletter={:},
  alsodigit={-}
}

\lstdefinelanguage{JavaScript}{
  morekeywords=[1]{break, continue, delete, else, for, function, if, in,
    new, return, this, typeof, var, void, while, with},
  morekeywords=[2]{false, null, true, boolean, number, undefined,
    Array, Boolean, Date, Math, Number, String, Object},
  morekeywords=[3]{eval, parseInt, parseFloat, escape, unescape},
  sensitive,
  morecomment=[s]{/*}{*/},
  morecomment=[l]//,
  morecomment=[s]{/**}{*/}, 
  morestring=[b]',
  morestring=[b]"
}[keywords, comments, strings]

\lstdefinelanguage[ECMAScript2015]{JavaScript}[]{JavaScript}{
  morekeywords=[1]{await, async, case, catch, class, const, default, do,
    enum, export, extends, finally, from, implements, import, instanceof,
    let, static, super, switch, throw, try},
  morestring=[b]` 
}

\lstalias[]{ES6}[ECMAScript2015]{JavaScript}

\lstdefinelanguage{HTML5}{
  language=html,
  sensitive=true,	
  alsoletter={<>=-},	
  morecomment=[s]{<!-}{-->},
  tag=[s],
  otherkeywords={
  >,
	<!DOCTYPE,
  </html, <html, <head, <title, </title, <style, </style, <link, </head, <meta, />,
	</body, <body,
	</div, <div, </div>, 
	</p, <p, </p>,
	</script, <script,
  <canvas, /canvas>, <svg, <rect, <animateTransform, </rect>, </svg>, <video, <source, <iframe, </iframe>, </video>, <image, </image>, <header, </header, <article, </article
  },
  ndkeywords={
  =,
  charset=, src=, id=, width=, height=, style=, type=, rel=, href=,
  fill=, attributeName=, begin=, dur=, from=, to=, poster=, controls=, x=, y=, repeatCount=, xlink:href=,
  margin:, padding:, background-image:, border:, top:, left:, position:, width:, height:, margin-top:, margin-bottom:, font-size:, line-height:,
  transform:, -moz-transform:, -webkit-transform:,
  animation:, -webkit-animation:,
  transition:,  transition-duration:, transition-property:, transition-timing-function:,
  }
}

\lstdefinestyle{htmlcssjs} {%
  basicstyle={\footnotesize\ttfamily},   
  frame=b,
  xleftmargin={0.75cm},
  numbers=left,
  stepnumber=1,
  captionpos=b,
  firstnumber=1,
  numberfirstline=true,	
  identifierstyle=\color{black},
  keywordstyle=\color{blue},
  ndkeywordstyle=\color{editorGreen},
  stringstyle=\color{editorOcher}\ttfamily,
  commentstyle=\color{brown}\ttfamily,
  language=HTML5,
  alsolanguage=ES6,
  alsodigit={.:;},	
  tabsize=2,
  showtabs=false,
  showspaces=false,
  showstringspaces=false,
  extendedchars=true,
  breaklines=true,
  literate=%
  {Ö}{{\"O}}1
  {Ä}{{\"A}}1
  {Ü}{{\"U}}1
  {ß}{{\ss}}1
  {ü}{{\"u}}1
  {ä}{{\"a}}1
  {ö}{{\"o}}1
}

\definecolor{mediumgray}{rgb}{0.3, 0.4, 0.4}
\definecolor{mediumblue}{rgb}{0.0, 0.0, 0.8}
\definecolor{forestgreen}{rgb}{0.13, 0.55, 0.13}
\definecolor{darkviolet}{rgb}{0.58, 0.0, 0.83}
\definecolor{royalblue}{rgb}{0.25, 0.41, 0.88}
\definecolor{crimson}{rgb}{0.86, 0.8, 0.24}

\lstdefinestyle{JSES6Base}{
  basicstyle={\footnotesize\ttfamily},   
  captionpos=b,
  columns=fullflexible,
  commentstyle=\color{mediumgray}\upshape,
  emph={},
  emphstyle=\color{crimson},
  extendedchars=true,  
  fontadjust=true,
  frame=b,
  identifierstyle=\color{black},
  keepspaces=true,
  keywordstyle=\color{blue},
  keywordstyle={[2]\color{darkviolet}},
  keywordstyle={[3]\color{royalblue}},
  ndkeywordstyle=\color{editorGreen},
  stringstyle=\color{editorOcher}\ttfamily,
  xleftmargin={0.75cm},
  numbers=left,
  stepnumber=1,
  firstnumber=1,
  numberfirstline=true,	
  rulecolor=\color{black},
  showlines=true,
  showspaces=false,
  showstringspaces=false,
  showtabs=false,
  stringstyle=\color{forestgreen},
  tabsize=2,
  title=\lstname,
  upquote=true  
}

\lstdefinestyle{JavaScript}{
  language=JavaScript,
  style=JSES6Base
}
\lstdefinestyle{ES6}{
  language=ES6,
  style=JSES6Base
}

%% file: table-markers.tex
\input{bite-circles-def.tex}

\usepackage{pifont} 
\usepackage{wasysym} 

\newcommand{\cmark}{\fullcircle}%
\newcommand{\xmark}{-}%
\newcommand{\xxmark}{\ding{53}}
\newcommand{\hmark}{\halfcircle}

%% file: bite-circles-def.tex
\newcommand{\circlesize}{1mm}
\newcommand{\smallcirclesize}{0.7mm}

\usepackage{tkz-euclide}

\tikzset{%
    pics/sema/.style args={#1/#2/#3}{code={%
        \tkzDefPoint(0,0){O}
        \tkzDrawSector[R,fill=#1](O,\circlesize)(90,90-#2)
        \tkzDrawSector[R,fill=#3](O,\circlesize)(90-#2,90-360)
    }},
}

\tikzset{%
    pics/smallfullcircle/.style args={#1/#2/#3}{code={%
        \tkzDefPoint(0,0){O}
        \tkzDrawSector[R,fill=#1](O,\smallcirclesize)(90,90-#2)
        \tkzDrawSector[R,fill=#3](O,\smallcirclesize)(90-#2,90-360)
    }},
}

\newcommand{\fullcircle}{
\tikz\pic{sema=black/180/black};
}
\newcommand{\halfcircle}{
\tikz\pic{sema=black/180/white};
}
\newcommand{\emptycircle}{
\tikz\draw (0,0) circle (\circlesize);
}

%% file: sections/intro/v2.tex
\section{Introduction}

A typical web user is required to maintain many passwords for their online accounts, a task that requires unreasonable cognitive burden. 
Password managers are widely recommended by the experts to relieve users of such burden, and if designed well, bring extra security and usability benefits through the use of strong passwords and streamlining the authentication process, respectively.

HTML Autofill~\cite{noauthor_html_nodate-1} is the most widely-used framework for password managers
to assist browser users during authentication. Here, the password manager interprets the web page, based on HTML elements and attributes, to determine which input fields can be automatically populated on behalf of the user, and if so, what type of information is expected; e.g., a password or a credit card number.
An (in-browser) password manager can then offer appropriate stored user secrets accordingly to automatically populate the fields.
Unfortunately, HTML Autofill is prone to behaving incorrectly since it is essentially a heuristic approach based on educated guesses to interpret the front-end markup language~\cite{stajano_password-manager_2014}. 
To make matters worse, with the growing prevalence of \emph{single-page applications} (SAPs), HTML Autofill is proving ineffective in correctly populating credentials. 
This is because SAPs often leverage the JavaScript Fetch API~\cite{noauthor_using_nodate} instead of HTML form submission~\cite{noauthor_html_nodate}, and username and password fields are no longer necessarily coupled under a form element, forcing password managers to implement additional best-effort heuristics. 
Such issues highlight a need for specialised solutions to provide programmatic management of passwords and other user secrets. 

W3C's Credential Management API~\cite{west_credential_2019} is an existing solution in this regard. 
The API provides a simple mechanism for web applications to store and retrieve user credentials in browser storage.
It enables programmatic access to user credentials.
However, Credential Management API is only available to so-called \emph{native} user agents, i.e., in-built browser password managers, and cannot be used to store and retrieve credentials into and from password manager \emph{extensions}.

In\ 2020, Stobert et al.\ proposed a remodelled password manager, ByPass, that communicates directly with the web application's back-end through a bespoke API~\cite{stobert_bypass_2020}. 
Similar to Credential Management API, ByPass also eliminates issues caused by misinterpretation by introducing a programming interface.
Furthermore, ByPass is not susceptible to credential theft by front-end threats such as cross-site scripting, and is able to provide enhanced services such as account deletion and password renewal. 
Despite all its benefits, ByPass radically transforms the user experience. 
The user is expected to launch the password manager and select a website to initiate the login process, instead of navigating to the website through the browser as is conventional. 
Consequently, Bypass not only takes away the control that web app developers relish today over the login user experience, it also requires users to develop and employ a new mental model.

Password managers are tasked with handling user credentials that are inherently sensitive. 
Hence, any password management governing framework must be evaluated based not only on its functionality features but also on whether it provides security services that help protect against credential theft attacks. 
Two prominent categories of such attacks are code injection and man-in-the-middle (MitM) attacks. 
Credentials can be stolen if client-side scripts, e.g., JavaScript code, is successfully injected on the client side by malicious entities external to the browser or by browser extensions. 
The former is the well-known \emph{cross-site scripting} (XSS) attack, and we denote the latter by the term \emph{extension code-injection} (ECI). 
On the other hand, while general MitM attacks are instigated by external network entities, a more subtle version may occur as a result of faulty or compromised TLS proxies~\cite{oneill_tls_2014}. 
We call this a \emph{TLS-proxy-in-the-middle} (TPitM) attack. 
While \emph{Content Security Policy} (CSP)~\cite{csp} can mitigate against XSS attacks, it is not effective against ECI attacks, and although TLS can defeat general MitM attacks, it does not provide any protection against TPitM attacks. 
HTML Autofill or Credential Management API do not provide any security services that can help mitigate against ECI or TPitM attacks. 
ByPass's architecture makes it intrinsically secure against XSS and ECI attacks as the client-side is \say{bypassed}, however there is no mitigation against TPitM attacks. 

\paragraph{Our Contributions.}
In this paper, we propose \emph{\protoname}, a web governance framework that mediates between web applications and password managers to orchestrate programmable registration and authentication sessions. 
Crucially, \protoname\ is positioned between the web application front-end and the password manager client, operating natively in the browser. 
This architectural choice enables \protoname\ to provide programmatic password management services to both native and extension password managers while preserving developer user experience control and user mental models. 
We design two APIs for web applications and password managers to communicate with \protoname, respectively. 
Vital security services such as web application authentication and authenticated key exchange between web applications and password managers are built into the \protoname\ APIs, which allow application-level \emph{end-to-end encryption} (E2EE) to be set up between the password manager and the web application back-end. 
This provides an effective security mitigation mechanism against XSS, ECI, and TPitM attacks. 

Apart from the main functional and security services discussed above, \protoname's design offers extra benefits. 
By authenticating web applications, \protoname\ is able to facilitate and enable password managers to rely on more accurate \emph{web application based credential mapping}, which aligns with the distributed nature of web applications and avoids the issues with domain-based mapping~\cite{carr_revisiting_2020,oesch_emperors_2021,huaman_they_2021,blanchou_password_2013}.
Furthermore, \protoname\ resolves the race condition issues when \emph{multiple password managers} are in use by harmonising password manager registration and selection prompts. 
We have implemented \protoname\ in Mozilla Firefox. 
All project artefacts, including the code, are available at \url{https://github.com/alichry/berytus}. 

The remainder of the paper is structured as follows:
We cover the related work in Section~\ref{sec:bg}.
The framework architecture is discussed in Section~\ref{sec:prp-fwk}. 
Security and functionality evaluations are given in Sections~\ref{sec:sec-eval} and \ref{sec:fun-eval}, respectively, conclusions in Section~\ref{sec:fw}, and further information on our implementation in Appendix~\ref{sec:implementation}.

%% file: sections/background/v2.tex
\section{Background and Related Work}
\label{sec:bg}

We give an overview of frameworks governing password managers and their security and functionality properties.
Since password managers store various types of credentials besides passwords, we use the term \emph{secret manager} henceforth.

\subsection{Existing Frameworks}

We give further details on the inner workings of HTML Autofill, Credential Management API, and ByPass. 
Figure~\ref{fig:background} provides a high-level comparison of the architectures of these major frameworks (as well as a preview of that of \protoname).

\paragraph{HTML Autofill: A UI-based Heuristic.}
Here, HTML input fields are filled by the browser (technically, the \say{user agent}) with relevant data, e.g., personal information and secrets, on behalf of the user. 
The filled data is either generated on the fly (e.g.\ a proposed password) or retrieved from what has been captured and stored during an earlier browsing session. 
The HTML Standard outlines the autofill guidelines for user agents~\cite{noauthor_html_nodate}, however, secret manager extensions are at liberty to provide the autofill functionality. 
Web apps can support the filling process by integrating the HTML \texttt{autocomplete} attribute into relevant input fields. 
This aspect of the HTML Standard might not be implemented for some login forms, leaving secret managers to rely on ad hoc heuristics for input field classification to determine which input fields to fill. 
This leads to interaction issues between web apps and secret managers causing inconvenience for users, e.g., the absence of input \emph{hints} hinders the filling process~\cite{huaman_they_2021}.
Therefore, while Autofill is highly \emph{deployable}~\cite{bonneau_quest_2012}, it is prone to behaving erratically due to imperfect and varying heuristics.
Besides, Autofill is \emph{forceful}: web apps cannot officially disable its behaviour. 
The HTML Standard concurrently specifies a method for web apps to disable Autofill (by setting \texttt{autocomplete} to \texttt{off}) and a suggestion for user agents to ignore such a declaration at their discretion~\cite{noauthor_html_nodate}.

\paragraph{Password-Manager Friendly: An Autofill Extension.}
Motivated by the lack of required declarative hints for input field classification in HTML Autofill, Stajano et al.\ proposed Password-Manager Friendly (PMF), an additional set of HTML semantic labels to ensure correct secret management behaviour~\cite{stajano_password-manager_2014}. 
Unlike the HTML Autofill, PMF aids secret managers in detecting submission errors and different form types, including login, registration, password reset, and password change. 
If web apps incorporate those additional semantics into their forms, secret managers would no longer need to rely heavily on heuristics and this would lead to the reduction of interaction issues in HTML Autofill. 

\paragraph{Credential Management API: A Credential Storage API.}
As Autofill was designed for user agents to aid users in \emph{HTML forms}, it became difficult to detect sign-in ceremonies leveraging the Fetch API~\cite{west_credential_2019,noauthor_using_nodate}.
When JavaScript Fetch API is used, credential submission over HTTP is not necessarily tied to an HTML form, making it troublesome for Autofill heuristics to detect username and password fields since they are now separated. 
As a result, secret managers may fail to fill and save passwords. 
Besides, user agents lacked support for federated sign-ins in HTML Autofill, and password change could be further supported by requiring web apps to notify user agents when credentials have been changed. 
Credential Management API was proposed to ensure improved credential management and to support users with federated sign-ins~\cite{west_credential_2019}. 
Fundamentally, it offers a programming interface for web apps to store and retrieve credentials into and from the user agent. 
Using JavaScript, a web app can insert a \texttt{PasswordCredential}, consisting of a username and a password, into the browser storage.
When the user visits the web app in the future, the web app can programmatically retrieve the password credential from the browser storage.
\emph{Web Authentication} (WebAuthn)~\cite{balfanz_web_2019} is an extension to Credential Management API that enables digital signature based authentication by introducing the new {\tt PublicKeyCredential}. Currently, some secret manager extensions act as third-party public-key credential providers by intercepting the WebAuthn API calls~\cite{noauthor_terms_nodate}. 

\paragraph{ByPass: A Secret Manager -- Website Back-end Interface.}
Motivated to address the usability issues of secret managers, Stobert et al.\ proposed a re-imagined secret management model where their manager, ByPass, communicates directly with the web app's back-end \cite{stobert_bypass_2020}.
As ByPass directly interacts with the web app's back-end, front-end security vulnerabilities such as XSS are eliminated from the attack surface.
To log into a website, the user needs to open ByPass and search for the website they wish to visit within ByPass's UI, instead of navigating to the website through the browser. 
ByPass then handles the account registration or authentication process under its own UI and not the web app's, and only if the process succeeds, ByPass opens the website homepage. 
Fundamentally, all account-related operations such as password change or account deletion can also be carried out by the secret manager and the user only needs to interact with the manager's interface. 
As a result, users are now required to develop a different mental model and web apps lose control over their login user interfaces.

\begin{figure*}[t]
\centering
\includesvg[width=\linewidth,inkscapeformat=eps]{fig1-as-4/all-in-one-complete.svg}
\caption{Comparing HTML Autofill, Cred.\ Mgmt API, ByPass, and \protoname\ architecture}
\label{fig:background}
\end{figure*}

\subsection{Security Threats and Mitigations} 
\label{sec:security_issues}
The above governance frameworks facilitate the communications between secret managers and web apps. 
Two prominent security threats to such communications are code injection and MitM attacks, discussed here in more detail. 

\paragraph{Code Injection: XSS and ECI.}

Both HTML Autofill and Credential Management API are frameworks that require secret managers to communicate credentials in plaintext with the web app front-end.
By design, these frameworks leave the user's credential accessible to JavaScript code in the web page. 
Therefore, an adversary that is able to inject such code is able to steal the user's credential~\cite{silver_password_2014,stock_protecting_2014,west_credential_2019,oesch_that_2020}.
Such code injection can occur through cross-site scripting (XSS) or by a malicious extension, an attack that we call extension code-injection (ECI). 
\emph{Content Security Policy} (CSP)~\cite{csp} is a standardised web security policy designed to mitigate against XSS, but not ECI. 
Alternatively, in 2014, Stock and Johns~\cite{stock_protecting_2014} proposed a \emph{credential tokenisation} mechanism for secret managers conducting HTML Autofill which mitigates against credential theft by malicious JavaScript code.
The credentials are provided on the HTML document as \emph{tokens} and substituted with the genuine credentials in the dispatched HTTP request payload using a search and replace algorithm. 
Tokenisation ensures that the credentials are not available in the clear to the front-end, and by extension any front-end eavesdropping adversaries, including cross-site scripting attackers and malicious browser extensions. 
In 2020, Oesch and Ruoti hunted for XSS-safe secret managers implementing this credential tokenisation mechanism and found that none did~\cite{oesch_that_2020}.
They justify such absence based on the limitations imposed by current browsers on extensions, disallowing them from manipulating HTTP request bodies.
In general, tokenisation assumes that passwords are communicated without any transformations such as hashing applied to them, which is at odds with modern recommendations for password treatment such as dedicated password hashing algorithms (see e.g.,~\cite{pwd-hash-comp}) and is incompatible with any other password-based protocol (e.g., SRP~\cite{wu_secure_1998}) that the client-side scripts may wish to execute. 
This may well be the reason that none of the native secret managers in Chrome, Safari, or Firefox implement tokenisation either. 

\paragraph{MitM and TPitM.}

Credentials travel from the user's browser to the web app's back-end, and hence are at the risk of credential theft through MitM network interception.
As is widely known, TLS renders MitM interception futile as the transmitted data is encrypted.
However, TLS proxies are capable of decrypting HTTP requests before relaying them to the final destination. 
Hence, a subverted TLS proxy in the middle (TPitM) enables an adversary to steal the transmitted credentials~\cite{oneill_tls_2014}. 
TLS proxies are routinely deployed in IT-managed environments, e.g., in corporate offices, and thus TPitM poses a security threat to corporate users. 
Unfortunately, none of the frameworks support a mitigation strategy to combat credential theft through TPitM.

\subsection{Functionality and Security Comparison}


\begin{table}[t]
    \newcommand{\provides}{\cmark}
    \newcommand{\lacks}{\xmark\ }
    \setlength{\tabcolsep}{3pt}
    \centering
    \caption{
    Comparison of existing frameworks based on selected functional properties (1st group), security properties (2nd group), and extra desirable properties (3rd group)
    }
    \label{table:existing_gaps}
    \begin{tabular}{lccc}
        \toprule
        Property & Autofill & CredMgmt & ByPass \\ 
        \midrule
        Provides programmatic secret management & \lacks{} & \provides{} & \provides{} \\
        Compatible with extension secret managers & \provides{} & \lacks{} & \provides{} \\
        Preserves UX control \& user mental model & \provides{} & \provides{} & \lacks{} \\
        \midrule
        Enforces uniform baseline security policies & \lacks{} & \provides{} & \provides{} \\
        Provides mitigation against code injection & \lacks{} & \lacks{} & \provides{} \\
        Provides further mitigation against MitM & \lacks{} & \lacks{} & \lacks{} \\
        \midrule
        Enables web app based credential mapping & \lacks{} & \lacks{} & \lacks{} \\
        Supports signature based authentication & \lacks{} & \provides{} & \lacks{} \\
        Supports credential customisation & \lacks{} & \lacks{} & \lacks{} \\
        \bottomrule
    \end{tabular}
    \vskip 0.25em
    \provides{}: provides property, \quad
    \lacks{}: lacks property
\end{table}

Here, we give a comparison between the existing frameworks in terms of functionality and security, and provide an overview of the properties \protoname\ aims to achieve. 
The discussed properties are summarised in Table~\ref{table:existing_gaps}. 

From the functionality point of view, we consider three main properties. 
Firstly, as we have seen, a programmable interface allows programmatic management of secrets, which in turn makes web app and secret manager behaviours predictable and eliminates the need to use heuristics. 
Only Credential Management API and ByPass provide such programmable interface. 
Secondly, an open framework must not only support native secret managers but also extension secret managers. 
Credential Management API is not currently available to extensions, but HTML Autofill and ByPass are agnostic to whether the secret manager is native or an extension. 
Finally, any framework should ideally be compatible with established user mental models and leave the control over the UX to the web app developers. 
HTML Autofill and Credential Management API do this, but ByPass radically changes the UX. 

We consider three security services the frameworks can provide. 
Firstly, any security-conscious framework must enforce certain baseline security policies such as only allowing credentials on HTTPS web apps. 
Otherwise, credentials would be exposed to simple MitM adversaries. 
Credential Management API and ByPass do this. 
Secondly, mitigation against code injection is desirable as CSP is not effective against ECI and tokenisation is unlikely to gain any popularity. 
ByPass eliminates such threats through \say{bypassing} the front-end, but HTML Autofill and Credential Management API do not provide any mitigation. 
Finally, providing further mitigation against more sophisticated MitM attacks such as TPitM would be a bonus that unfortunately no current framework offers. 

We also consider three extra desirable properties. 
Firstly, all existing frameworks employ \emph{domain-based} credential mapping, which could be problematic for a web app residing under different domains, or for multiple web apps residing under the same domain, as Huaman et al. demonstrate~\cite{huaman_they_2021}. 
A more accurate mapping strategy would be based on individually identified and authenticated \emph{web apps} rather than domains.
Secondly, support for automated forms of authentication such as authentication based on digital signatures, as provided by Credential Management API (WebAuthn), is a further desirable property. 
Finally, none of the existing frameworks are designed to support custom credential structures, such as those with multiple usernames or multiple passwords, which could be an impediment for less conventional web apps.

As Table~\ref{table:existing_gaps} shows, none of the existing platforms provides a good coverage of the main functional and security properties, let alone the extra desirable ones. 
The design aim for \protoname\ is to provide all these properties.

%% file: sections/method/v2.tex
\section{Proposed Governance Framework}
\label{sec:prp-fwk}

\protoname\ is designed as a governance framework for programmable account registration and authentication sessions through secret managers. 
\protoname{} offers two integration pathways for web apps: 
base integration requiring front-end changes only, and 
full integration with enhanced security requiring front-end and back-end changes.
Here, we will discuss the architecture and operational design.

\subsection{Architectural Overview}
\label{sec:architectural_overview}
As an orchestrator, \protoname{} operates natively in the browser, sitting between the web app front-end and the secret manager client. 
\protoname\ introduces two APIs, a Web API~\cite{noauthor_web_2023} for web apps and a WebExtensions API~\cite{swan_webextensions_2016} for secret managers.  Essentially, \protoname{} relays the instructions given by the web app via the Web API to the secret manager via the WebExtensions API. 
In this section, we unpack the components, routines, and facilities of \protoname{}.

\subsubsection{Components.}
\label{sec:architectural_overview_components}

The building blocks of \protoname\ are shown in Figure~\ref{fig:overview}.
We conceptualise account-related processes, e.g., authentication or registration, as \emph{operations}, each a series of one or more actions, resembling a form with multiple steps. 
Furthermore, we introduce the concept of a \emph{channel} to reflect an active logical link between the web app and the secret manager. 
The channel holds the two \emph{actor} objects, one for the web app and one for the secret manager, containing identifying information of each party, allowing the two sides to identify each other. 
\protoname\ also provides routines for authenticating web apps as explained later. 
Web apps can rely on the browser's attestation to trust the declared identity of the secret managers. 
Mutual identification and authentication are important for web apps such as internet banking which may wish to set a security policy that restrict their interaction to specific secret managers, and for secret managers to locate the corresponding account records in their databases.

\begin{figure*}[t]
\centering
\includesvg[width=\textwidth]{svgs/berytus-overview.svg}
\caption[
Illustration of the \protoname{} communication model and components.
]{
Illustration of the \protoname{} communication model
between the web application and the secret manager along with their components.
}
\label{fig:overview}
\end{figure*}

There are two actor specialisations. The first specialisation is the \emph{\MakeLowercase{\originactor{}}} and is exclusive for web apps. 
It reflects the web page's Uniform Resource Identifier (URI). 
The second specialisation, \emph{\MakeLowercase{\cryptoactor{}}}, can be used by both the web app and secret manager. 
It requires the backing of a (cryptographic) signing key.
A web app hosted at distinct resource locations will produce distinct \MakeLowercase{\originactor{}}s, one distinct actor for each distinct URI.
Conversely, if a web app uses a \MakeLowercase{\cryptoactor{}}, it will construct uniform actors across all of its resource locations. Similarly, if a secret manager creates a \MakeLowercase{\cryptoactor{}}, it will construct uniform actors across various desktop or mobile environments.

Finally, we appoint secret managers to construct a \emph{request handler} function, i.e., a callback function, to process the web app's instructions programmatically. 
This is invoked to process requests such as constructing an account password field. 
Taking this approach ensures that secret managers are afforded programmability. Programmability is a pivotal functionality, allowing secret managers to react dynamically, instead of merely behaving as a reactive credential provider, and to fulfill business requirements, e.g., sending an email notification once a credential is transferred to the web app. 

\subsubsection{Routines.}
\label{sec:architectural_overview_routines}

We describe the \protoname\ routines from a high-level perspective. 

\paragraph{Secret Manager Registration.}
To track registered and running secret managers, the \protoname{} WebExtensions API provides two primary methods, one for registration and one for de-registration, and both are available to an extension if it specifies the \texttt{\MakeLowercase{\protoname}} permission in its manifest file~\cite{noauthor_manifestjson_nodate}. The secret manager passes a request handler during registration and, subsequently, it gets stored in \protoname{}. The request handler is invoked when \protoname{} Web API calls are dispatched by the web app.  At a future point in time, the secret manager can de-register if needed and the stored request handler gets disposed.

\paragraph{Authenticating Web App Crypto Actors using Digital Certificates.}
In \protoname{}, web apps can be identified using \MakeLowercase{\originactor{}}s or \MakeLowercase{\cryptoactor{}}s. Origin actors are authenticated using the well-established certificate-based TLS website origin authentication. For web app \MakeLowercase{\cryptoactor{}}s, we propose a new X.509 v3 certificate extension: \protoname{} Signing Key Allowlist. This certificate extension specifies a list of one or more keys that can be used by the certificate’s subject and under which URIs each key can be used. Overall, at a logical level, the web app \MakeLowercase{\cryptoactor{}} public key should be certified by a trusted certificate authority. 
    
\paragraph{Prompting for Secret Manager Selection.} 
When the web app initiates the creation of a web app -- secret manager orchestration channel, the user is prompted through the browser UI to select a secret manager from the list of registered managers (see Figure~\ref{fig:berytus-scm-prompt}~(left) in the Appendix). 
Following the secret manager selection, the channel is created and one or more account-related operations, e.g., registration or authentication, can be initiated.
\paragraph{Mediating an Authenticated Key Exchange.}
\protoname{} mediates an authenticated Diffie--Hellman key exchange, specifically the elliptic curve-based X25519~\cite{bernstein_curve25519_2006}, using the web app's front-end as a medium to communicate cryptographic material passed between the web app's back-end and the secret manager. 
The exchanged keys are authenticated by each party's \MakeLowercase{\signingkey{}}. 

\subsubsection{Facilities.}
\label{sec:architectural_overview_facilities}

The above enable \protoname\ to offer the following facilities:

\paragraph{Unified Secret Management.} 
By requiring secret managers to register, the browser can keep track of running managers and users can select a secret manager on a per-session basis, ensuring coordinated support for \emph{multiple} secret managers.

\paragraph{Web App to Credential Mapping.}
Secret managers can leverage the web app actor's identifying material to distinguish between web apps and perform web app based credential mapping. 
The \MakeLowercase{\originactor{}} is used for website origin based identification which enables domain-based mapping, and the \MakeLowercase{\cryptoactor{}} is used for key-based identification which enables web app based mapping. 
The web app is at liberty to pick from the two actor specialisations as appropriate.

\paragraph{Application-level End-to-End Encryption.} 
Following an authenticated Diffie--Hellman key exchange, \protoname{} facilitates end-to-end encryption (E2EE) between the secret manager and the web app. 
The E2EE channel is set up at the \emph{application} level between the secret manager and the web app back-end. 
This provides a separate encrypted channel \emph{within} the network-level encrypted channel provided by TLS between the browser and the web server.
The web app back-end and the secret manager have exclusive access to the shared secret key for the application-level E2EE. 
To realise this, the web app back-end codebase needs to be changed to implement the necessary cryptographic functions.

\subsection{Operational Design}

We designed and implemented two operations: account creation for singing up, and account authentication for signing in. 
These two operations are generalised into the {\em login operation}.
In this section, we describe the operation approval pattern followed by an overview of the login operation, including intent, account creation, and account authentication.

\subsubsection{Operation Approval Pattern.}

Once an operation has been initiated by the web app, through interaction with the channel instance, the secret manager's request handler is invoked to approve the operation creation. The secret manager should resolve this request or reject it. 

\subsubsection{Login Operation Intent.}

Generally, login forms prompt the user for their intent, i.e., to authenticate using an existing account, or to create a new account. 
We assume a similar design for the login operation as a branched operation dependent on user intent. 
The secret manager retrieves user intent and resolves the operation approval request as outlined in Figure~\ref{sqd:login_creation}. 
Alternatively, the web app can dictate the intent, 
e.g., in case the web app only allows authentication.

\subsubsection{Account Creation Operation.}
\label{sec:overview_registration} 

Inspired by how web apps typically conduct registration, we design the registration process using 
fields and user attributes.


\begin{table*}[t!]
    \newcommand{\produciblebyany}{app or manager}
    \newcommand{\produciblebyscm}{manager only}
    \newcommand{\produciblebywebapp}{app only}
    \setlength{\tabcolsep}{3pt}
    \centering
    \caption{%
    Overview of supported account field specialisations in \protoname{}
    }
    \label{table:different_field_types2}
    \begin{tabular}{lll}
        \toprule
        Field Specialisation & Sample value at registration & Value producible by \\ 
        \midrule
        Identity & identifier: \texttt{bob123} & \produciblebyany{} \\
        Foreign Identity & identifier: \texttt{bob@example.org} & \produciblebyany{}  \\
        Password & password: \texttt{RyU8HxsJjk332Dg} & \produciblebyany{} \\
        Secure Password {\footnotesize (SRP \cite{wu_secure_1998})} & salt: \texttt{\footnotesize 0edb53..} verifier: \texttt{\footnotesize 29c792..} & \produciblebyscm{} \\
        Key & public key: \texttt{\footnotesize 010101...} & \produciblebyscm{} \\
        Private Key & private key: \texttt{\footnotesize 010101...} & \produciblebyany{} \\
        \bottomrule
    \end{tabular}
\end{table*}

\paragraph{Credential as Fields.}
An account field resembles a single input field in a registration form.
A field typically represents either an identity value, e.g., a username, or a secret value, e.g., a password.
An account can have multiple fields, and thus a credential can be conceptualised as a set of fields. 
This design choice enables a high degree of flexibility for web apps in defining any combination of fields as an account record. 
Table~\ref{table:different_field_types2} lists the field specialisations \protoname{} supports.

\paragraph{Field Value Production.}
The main piece of information for each field, the field value, can be set by either the secret manager or the web app. 
Producing a field value can be done by prompting the user for an appropriate value, e.g., an email address, or by generating a conforming value, e.g., a password or a key. 
\protoname\ streamlines relevant field options, e.g., a password composition policy, to enable web apps to specify and secret managers to comply with validity requirements.
The decision to delegate field value production to the secret manager is made by the web app. 
However, as shown in Table~\ref{table:different_field_types2}, the web app cannot produce a field value for the Key field or Secure Password field as the field value's corresponding secret is not and should not be known to the web app.

\begin{figure}[t!]
\centering
\footnotesize
\begin{sequencediagram}
\renewcommand\unitfactor{0.5}
\tikzstyle{inststyle}+=[right=-0.8cm]
\newthread{webapp}{\webapp{}} 
\tikzstyle{inststyle}+=[right=0cm]
\newinst[1]{channel}{Channel}
\newinst[1]{op}{Operation}
\tikzstyle{inststyle}+=[right=0.8cm]
\newthread{sm}{\scm{}}

\postlevel

\begin{call}{webapp}{Login()}{channel}{\shortstack{
return \\ operation
}}
    \begin{call}{channel}{ApproveLoginRequest()}{sm}{return user intent}
    \end{call}
    \begin{call}{channel}{\textit{instantiate}}{op}{}
    \end{call}
\end{call}

\end{sequencediagram}

\caption[%
Sequence diagram of the \protoname{} login operation initiation process.
]{(Simplified) sequence diagram showing the interactions between the web app, channel, operation and the secret manager during the \protoname{} login operation initiation process. The relayed user intent is either an authentication intent or registration intent.}

\label{sqd:login_creation}

\end{figure}

\paragraph{Registering Fields.}
To register an account field, the web app must construct a specialised field object (\texttt{id}, \texttt{options}, \texttt{value}) and transmit it to the secret manager. 
The secret manager processes the received fields, produces field values for the fields with unspecified values, and returns them to the web app.

\paragraph{Rejecting and Revising Fields.}
A field value, whether produced by the secret manager or the user, can adhere to the value format but may be unusable due to app-specific constraints, e.g., username uniqueness.
Hence, for robustness, the operation enables the web app to reject a produced field value by the secret manager and \emph{request} a revision providing a rejection reason. 
Alternatively, the web app can \emph{propose} a usable field value as a revision to the secret manager.

\paragraph{Retrieving Identity Information.}
Apart from fields, web apps often request identity information such as name and address during registration. 
As such, the web app can retrieve the user's identity information, based on the OpenID Standard Claims~\cite[Section~5.1]{sakimura_openid_2014}, from the secret manager. 

\paragraph{Account Categorisation.}
Web apps can set a category on the account record. 
The account category is used as an arbitrary account type identifier, e.g., specifying a user role. 
This can be specified as an option in the account authentication operation to assist the user in selecting an appropriate account in cases where users may have multiple accounts with the same web app. 

\paragraph{Saving the Account and Transitioning to Account Authentication.}
To finalise the registration operation, the web app instructs the secret manager to save the account record into its database.
If successfully saved, the finalised account creation operation can be transformed into an account authentication operation
for the newly-created account record.

\subsubsection{Account Authentication Operation.}

The account authentication operation encompasses the steps involved in user authentication. 
At a minimum, a web app typically employs two challenges, an identification challenge and an authentication challenge. 
In the same spirit, we design and implement a {\em challenge-based} authentication paradigm in which authentication is composed of a number of challenge--response rounds carried out sequentially. 

\paragraph{Account Selection.}
To process an authentication operation, the secret manager must first assume a registered account. It is expected that the secret manager would prompt the user to select an account using its own UI facilities during the early phases of the operation, e.g., in the login operation approval process, 
the secret manager can prompt the user to select an account if any exists.

\paragraph{Challenge Communication Flow.}
The challenge is designed as a message passing interface: the web app sends a message and the secret manager responds to it with the expected data. This enables the implementation of multi-step protocols such as the Secure Remote Password (SRP) protocol~\cite{wu_secure_1998}.

\paragraph{Supported Challenges.}
\protoname\ is designed to support password authentication, digital signature-based authentication, SRP authentication, and off-channel one-time password authentication (e.g., by email or phone).
Furthermore, web apps can initiate custom authentication challenges backed by a messaging JSON schema to validate the communicated messages. 
Hence, an acquainted web app and secret manager pair can organise custom challenges at run time.

\paragraph{Approving the Challenge.}
Once a web app initiates a challenge, \protoname{} contacts the secret manager to approve the challenge initiation request. Following its approval, the web app can begin sending messages to the secret manager.

\paragraph{Closing or Aborting the Challenge.}
When the web app is satisfied by the response, it can close the challenge to imply successful completion. 
Otherwise, the web app can abort the challenge by providing an abortion reason code.

%% file: sections/security/v2.tex
\section{Security Evaluation}
\label{sec:sec-eval}

To evaluate the security aspects of \protoname{}, we discuss the security services it provides and how they help mitigate against prominent credential-theft attacks in comparison with existing frameworks. 

\subsection{Security Services}
\label{sec:security_services}
As an orchestrator, \protoname{} is able to provide security services to web apps and secret managers, including authentication of web applications and application-level end-to-end encryption. 
We discuss these services here in more detail.

\subsubsection{Certificate-based Web App Authentication.}
\protoname{} enables \emph{web app} authentication through certified key identifiers (see \MakeLowercase{\cryptoactor{}} in Section~\ref{sec:architectural_overview_components}). 
This is in contrast with using a certified domain name for \emph{website origin} authentication via TLS certificates.
This provides the following benefits.

    \paragraph{Accurate Credential Mapping.}
    Domain-based credential mapping, used currently by secret managers, has been shown to produce several issues including inaccurate mappings when  the same authentication system is used on multiple domains~\cite{huaman_they_2021}.
    Authenticating web apps rather than their website origin enables the more accurate \emph{web app based credential mapping} (see Facilities in Section~\ref{sec:architectural_overview_facilities}). 
    This is an origin-agnostic credential mapping approach that is better tailored to the distributed nature of web apps and addresses the
    identified need in the literature for \say{using credentials in multiple environments}~\cite{huaman_they_2021}. 
    \paragraph{Security Against Web App Impersonation.}
    Malicious web apps may impersonate other apps to phish user credentials.
    Both web app authentication and website origin authentication protect against such impersonation. However, the former is tailored towards web apps possibly deployed on multiple domains, while the latter is tailored towards websites solely hosted on distinct domains.

\begin{figure*}[t!]
\centering
\includesvg[width=0.9\linewidth,inkscapeformat=eps]{svgs/berytus-full-e2e.svg}
\caption{\protoname\ E2EE \& its effectiveness against TPitM and monkey-patching attacks}
\label{fig:berytus_full_e2e}
\end{figure*}

\subsubsection{Application-level End-to-End Encryption.}
As \protoname{} supports equipping web apps with cryptographic keys, it can facilitate an authenticated Diffie--Hellman key exchange between the secret manager and the web app back-end (see Routines in Section~\ref{sec:architectural_overview_routines}). 
This provides the following benefit.

    \paragraph{Encryption-based Security Against MitM Attacks.}
    Although network-level encryption provided by TLS can guard against general MitM attacks, it is ineffective against the more sophisticated TPitM attacks. 
    Application-level E2EE facilitated by \protoname\ establishes a secure (logical) channel between the secret manager and the web app back-end, providing an effective mitigation against credential theft by any adversaries on this path, regardless of their security privilege. 
    Figure~\ref{fig:berytus_full_e2e} clarifies how E2EE works in \protoname.

\subsection{Resistance Against Attacks}
In this section, we discuss relevant mitigation strategies in \protoname{} and other frameworks to combat prominent credential phishing and theft attacks.

\paragraph{Threat Model.}
There are three broad types of entities in the web user authentication and secret management ecosystem: web apps, the browser%
, and browser extensions (including secret managers). 
We assume that the browser's native code and its privileged execution context, including \protoname, are trusted. 
We also assume that secret managers are trusted entities, but web apps and other browser extensions could be malicious. 
We do not assume an extra-vigilant user, so the user may visit and not distinguish malicious websites and may install malicious browser extensions. 
However, note that if the user consents to an extension acting as their secret manager, the extension is given such permission and hence trusted.
Moreover, we assume that the user's secret manager will only transfer credentials under the same credential mapping strategy as that when it was stored, e.g., if the credential was registered under domain-based credential mapping, it will only be transferred when domain-based credential mapping is used. 
We also assume that certificates are trusted. 

\paragraph{Compared Frameworks.}
Here, we focus on comparing HTML Autofill and Credential Management API with our proposed framework, \protoname, as these three are all client-side frameworks and many of the security services we consider here are client-side services. 
ByPass uses a significantly different architecture (see Figure~\ref{fig:background}), bypassing the client-side, and is discussed towards the end. 

\paragraph{Credential Theft Attacks.}
We consider code injection attacks, specifically XSS and ECI, and MitM attacks, specifically by general non-privileged adversaries (denoted by MitM) and by privileged adversaries such as TLS proxies (denoted by TPitM). 
For code injection attacks, CSP is effective against XSS, but not against ECI, as CSP can prevent code injection from entities external to the browser, but extensions will still have the ability to inject JavaScript code in the execution context. 
CSP is available in all three frameworks and can be considered an orthogonal service for extra assurance. 
Credential tokenisation however, while effective against both XSS and ECI, is not available in any of the frameworks, and given its limitations, is not expected to be widely-deployed. 
For MitM attacks, network-level TLS encryption is effective against general MitM attacks, but not against TPitM attacks. 
The application-level E2EE service provided by \protoname\ on the other hand is effective both against code injection attacks and against TPitM attacks. 
E2EE ensures that only the secret manager and the web app back-end have access to plaintext credentials. 
This means that the web-app front-end (and hence any code injected into it) or any party in the middle (such as a TLS proxy) can only get their hands on encrypted credentials and both types of attacks are rendered ineffective. 
These discussions are summarised in Table~\ref{tab:cred-theft}. 

\newcommand{\feff}{\cmark}
\begin{table}[t]
    \centering
    \caption{Selected security services, frameworks on which they are available (out of HTML Autofill, Cred.\ Mgmt.\ API, and \protoname), and security service effectiveness against credential-theft attacks (XSS, ECI, MitM, and TPitM)}
    \label{tab:cred-theft}
    \begin{tabular}{l@{\quad\quad}l@{\quad\quad}c@{\quad}c@{\quad}c@{\quad}c}
        \toprule
        Security Service & Framework & ~XSS~ & ~ECI~ & MitM & TPitM \\
        \midrule
        Content Security Policy & all & \feff & & & \\
        Credential Tokenisation & none & \feff & \feff & & \\
        Network-level TLS encryption & all & & & \feff & \\
        Application-level E2EE & \protoname & \feff & \feff & \feff & \feff \\
        \bottomrule
    \end{tabular}
    \vskip 0.25em
    \feff\!: credential theft via given attack is mitigated
\end{table}

\paragraph{Credential Phishing Attacks.}
In these attacks, users are lured into malicious websites that resemble other legitimate ones in the hope to steal the user's credentials for the legitimate websites. 
Web app origin authentication, via TLS certificates, along with domain-based credential mapping can effectively protect users against such attacks and are widely used by secret managers across all frameworks. 
\protoname\ provides a further alternative mitigation in this regard via the combination of app authentication and app-based credential mapping. 
As discussed before, in certain scenarios where there is no clear one-to-one mapping between web apps and domains, e.g., same apps on multiple domains, or multiple apps on the same domain, app-based mapping may be more appropriate and hence more accurate. 
Secret managers are free to choose the best mapping strategy in \protoname\ for the web apps with which they are communicating. 

\paragraph{ByPass.}
As in ByPass, secret managers directly communicate with web app back-ends, the code injection threat is eliminated as no login webpage is rendered. 
TLS is still effective against general MitM attacks, however a TLS Proxy will still be able to decrypt the communications between the secret manager and the web app back-end and since ByPass does not provide any further protection in this regard, it remains vulnerable to TPitM attacks. 